\documentclass[journal]{IEEEtran}
\newcommand{\RNum}[1]{\uppercase\expandafter{\romannumeral #1\relax}}
\usepackage{mathrsfs}
\usepackage{amssymb}
\usepackage[mathcal]{euscript}
\usepackage{diagbox}
\usepackage{cite}
\usepackage[T1]{fontenc}
\usepackage{graphicx}
\usepackage{CJKutf8}
\usepackage{makecell}
\usepackage{psfrag}
\usepackage{url}
\usepackage{stfloats}
\usepackage{amsmath}
\usepackage{array}
\usepackage{float}
\usepackage{multirow} 
\usepackage{hyperref}
\usepackage{amsthm}
\usepackage{color}
\usepackage{multicol}
\usepackage{stfloats}
\usepackage{enumerate}
\usepackage{subfigure}
\usepackage{booktabs}  

\allowdisplaybreaks[4]
\usepackage[ruled]{algorithm2e}
\usepackage{algorithmic} 
\ifCLASSINFOpdf
\else
\fi
\begin{document}
\begin{CJK}{UTF8}{gbsn}
%
\title{Generative AI for Space-Air-Ground Integrated Networks}



%
	\author{Ruichen Zhang,  Hongyang Du, Dusit Niyato,~\IEEEmembership{Fellow,~IEEE}, Jiawen Kang, Zehui Xiong, \\ Abbas Jamalipour,~\IEEEmembership{Fellow,~IEEE},   Ping Zhang,~\IEEEmembership{Fellow,~IEEE}, and Dong In Kim,~\IEEEmembership{Fellow,~IEEE}

\thanks{R. Zhang and D. Niyato are with the College of Computing and Data Science, Nanyang Technological University, Singapore (e-mail: ruichen.zhang@ntu.edu.sg, dniyato@ntu.edu.sg).}

\thanks{H. Du is with the Department of Electrical and Electronic Engineering, University of Hong Kong, Pok Fu Lam, Hong Kong (e-mail: duhy@eee.hku.hk).}

\thanks{J. Kang is with the School of Automation, Guangdong University of Technology, China (e-mail: kavinkang@gdut.edu.cn).}

\thanks{Z. Xiong is with the Pillar of Information Systems Technology and Design, Singapore University of Technology and Design, Singapore (e-mail: zehui\_xiong@sutd.edu.sg).}

\thanks{A. Jamalipour is with The University of Sydney, Sydney NSW 2006, Australia (e-mail: a.jamalipour@ieee.org).}

\thanks{P. Zhang is with the State Key Laboratory of Networking and Switching Technology, Beijing University of Posts and
Telecommunications, Beijing 100876, China (e-mail: pzhang@bupt.edu.cn).}

\thanks{D. I. Kim is with the Department of Electrical and Computer Engineering, Sungkyunkwan University, Suwon 16419, South Korea (email: dongin@skku.edu).}

}
\maketitle

\begin{abstract}
Recently, generative AI technologies have emerged as a significant advancement in artificial intelligence field, renowned for their language and image generation capabilities.  Meantime, space-air-ground integrated network (SAGIN) is an integral part of future B5G/6G for achieving ubiquitous connectivity. Inspired by this, this article explores an integration of generative AI in SAGIN, focusing on potential applications and case study. We first provide a comprehensive review of SAGIN and generative AI models, highlighting their capabilities and opportunities of their integration. Benefiting from generative AI's ability to generate useful data and facilitate advanced decision-making processes, it can be applied to various scenarios of SAGIN. Accordingly, we present a brief survey on their integration, including channel modeling and channel state information (CSI) estimation, joint air-space-ground resource allocation, intelligent network deployment, semantic communications, image extraction and processing, security and privacy  enhancement. Next, we propose a framework that utilizes a generative diffusion model (GDM) to construct channel information map to enhance quality of service for SAGIN. Simulation results demonstrate the effectiveness of the proposed framework. Finally, we discuss potential research directions for generative AI-enabled SAGIN.
\end{abstract}


\section{Introduction}

With the development of wireless communications, the goal of future  6th generation (6G) mobile networks is to achieve comprehensive connectivity with high performance and reliability. To meet the requirements, the space-air-ground integrated network (SAGIN) has been proposed, which combines ground, air, and space communications in which their functions efficiently complement each other for seamless and high-speed connections \cite{8368236}. Specifically, the space-based network is composed of the geostationary-earth orbit (GEO) satellites and the low-earth orbit (LEO) satellites, which can provide global coverage, high-speed data transmission, and ensure reliable communication even in remote or under-served regions. For example, Starlink, a massive satellite network developed by SpaceX, is a constellation of thousands of LEO satellites designed to provide broadband internet service across the globe\footnote{https://www.space.com/spacex-starlink-satellites.html}. The air-based network is composed of the aircraft, unmanned aerial vehicles (UAVs) and high altitude platforms (HAPs), which can swiftly respond to fluctuating demand, and provide coverage in areas where ground or space-based solutions might be less effective. The ground-based network is composed of cellular towers, small cells, and edge computing facilities. They enable the ultra-low latency communication and support dense urban connectivity. Thanks to its integrated multi-layered approach, SAGIN is able to provide a seamless, adaptive, and resilient communication infrastructure that meets the diverse needs of users across varied environments.

With the development of artificial intelligence (AI) technologies, generative AI has attracted more and more attention in terms of research, development, and applications. In particular, generative AI describes methods or instances that can be used to create new content, including audio, codes, images, texts, simulations, and videos \cite{du2024enhancing}. For example,  DALL-E~2\footnote{https://openai.com/dall-e-2}, a generative AI system from open AI, showcases the prowess of generative AI by synthesizing novel images based on text descriptions (i.e., called prompts). Regarding generative AI, key models include generative adversarial networks (GANs), diffusion models, and variational autoencoders (VAEs), all of which have found practical applications in real-world scenarios. For example, GauGAN\footnote{https://blogs.nvidia.com/blog/2019/03/18/gaugan-photorealistic-landscapes-nvidia-research/}, an NVIDIA AI demonstration for generating photo-realistic images, enables individuals to leverage GAN to craft landscapes.  {Also, generative AI facilitates dynamic resource allocation by creating detailed simulations and generating relevant data. For example, generative diffusion models (GDMs) utilize generative models to enhance decision-making processes by simulating various network states and conditions. The diffusion agent can use different strategies for different scenarios. By continuously generating new data and updating its strategies, the DBM-based method can adapt to changing network conditions, achieving efficient and dynamic resource management \cite{du2024enhancing}. In terms of data usage, generative AI can operate with both available and newly generated data. With existing data, generative AI models can analyze historical patterns to inform future resource allocation strategies \cite{xu2023unleashing}.}

Unlike traditional discriminative AI models that focus primarily on analyzing and processing existing data, generative AI holds significant advantages:
\begin{itemize}
\item {\bf Content Generation:} Generative models are capable of creating entirely new, high-quality content that is contextually relevant, thereby reducing the necessity for exhaustive data collection and manual content creation.
\item {\bf Flexibility and Adaptability:} Generative AI can adapt and learn from a diverse array of unstructured data sources, rendering it more versatile for handling a variety of applications.
\item {\bf Enhanced Simulation and Optimization Capabilities:} Through simulating intricate scenarios or systems, generative AI facilitates better forecasting, planning, and decision-making processes.
\end{itemize}
{While generative AI is primarily known for content generation, its ability to create strategic data and policies sets it apart from traditional AI approaches. Generative AI can produce data-driven strategies for resource allocation, which are continuously refined based on real-time feedback and changing network conditions. This method enhances adaptability and optimization capabilities compared to traditional AI approaches. For instance, traditional AI might rely on static rules or predefined models, while generative AI continuously evolves by generating new data and learning from it. }

\begin{table*}[!t]
\caption{Summary of typical generative AI models in SAGIN.} 
\includegraphics[width=\textwidth]{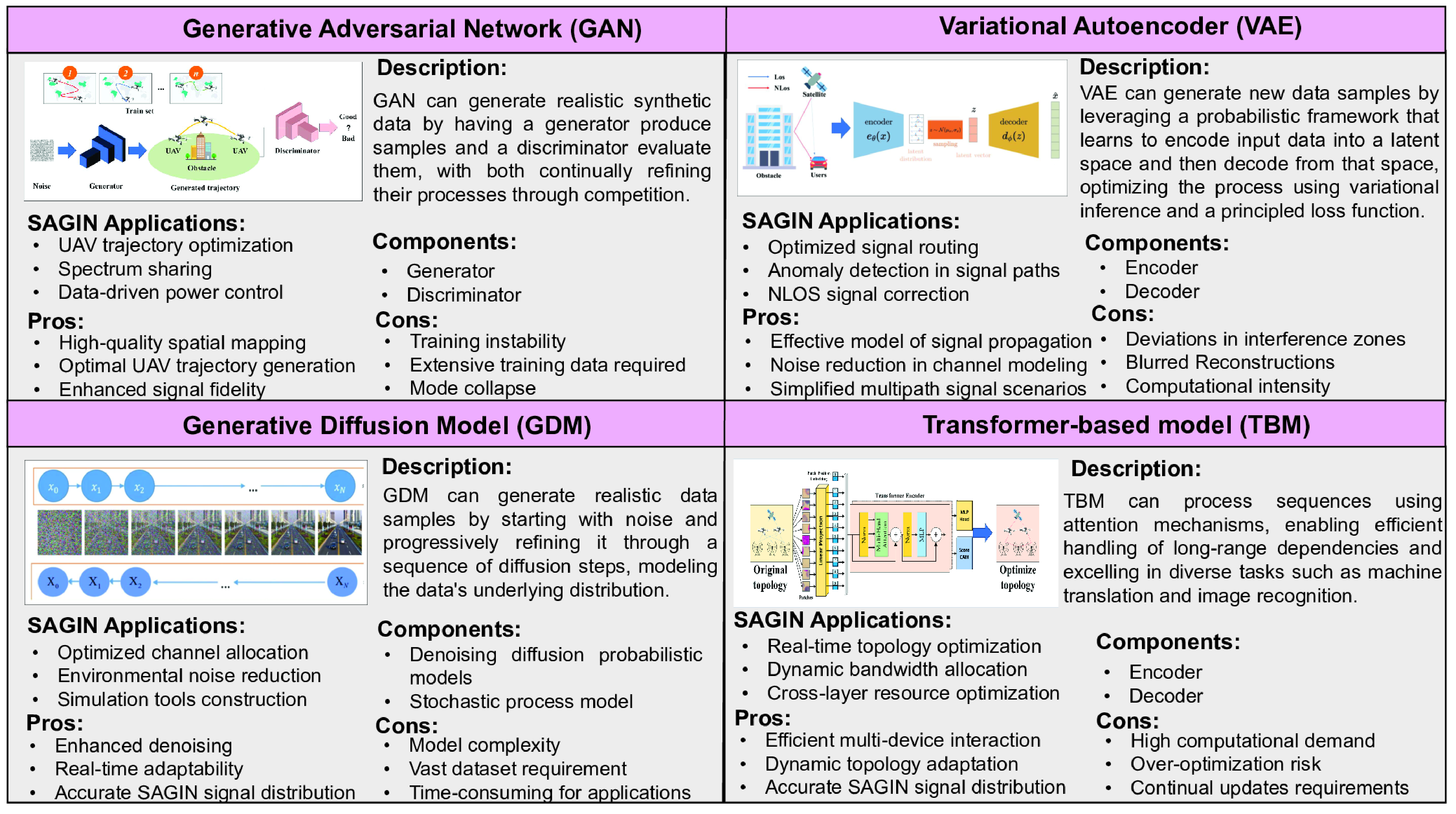}
\label{table_generative AI}
\end{table*}

{Integrating generative AI with SAGIN is expected to significantly transform wireless communication networks and digital content creation. By leveraging the comprehensive connectivity provided by SAGIN, generative AI facilitates the seamless generation and dissemination of high-quality information across disparate networks, such as real-time traffic updates, emergency alerts, and personalized service recommendations \cite{du2024enhancing}. For example, during emergencies, generative AI can create and distribute critical information, such as evacuation routes and emergency contacts, ensuring that individuals and authorities are fully informed and prepared to respond. Generative AI's adaptability can also be used for intelligent network management across all layers of SAGIN. By analyzing network traffic, user requirements, and environmental conditions, generative AI can dynamically allocate resources, predict network congestion, and optimize various space, air, and ground network performances. This convergence enhances user experience by delivering personalized services, real-time information, and interactive content through SAGIN's extensive network infrastructure. Furthermore, the ability of generative AI to create realistic simulations of network scenarios allows for proactive maintenance and optimization, reducing downtime and improving overall network efficiency.} Although the integration of generative AI and SAGIN has several potential advantages, there are still the following issues that need to be addressed:
\begin{itemize}
\item {\bf Q1:} Which issues in SAGIN can be solved with generative AI? 
\item {\bf Q2:} What kind of generative AI can be used for SAGIN?
\item {\bf Q3:} How to apply generative AI in SAGIN?
\end{itemize}

Motivated by these, this article aims to provide forward-looking research to address the aforementioned questions, with the contributions summarized as follows:

\begin{itemize}
\item {\bf First:} We provide an overview of the SAGIN and introduce different types of generative AI techniques. {\textit{To the best of the authors' knowledge, there is currently no review or magazine article that examines the holistic integration of generative AI within the SAGIN frameworks.}}
\item {\bf Second:} We explore the existing issues and solutions related to generative AI-enabled SAGIN. Additionally, we discuss the challenges associated with implementing generative AI to improve various aspects of SAGIN.
\item {\bf Third:} To enhance the quality of service for SAGIN, we construct a novel generative AI-enabled channel information map. Here, the GDM is adopted to perform channel classification. Simulation results validate the effectiveness of the proposed framework.
\end{itemize}

\section{OVERVIEW OF SAGIN AND GENERATIVE AI }

In this section, we first review SAGIN toward its openness to leverage the potentials of generative AI. Then, typical  generative AI models are introduced.

\subsection{Overview of SAGIN}
SAGIN network represents a cross-layer architecture in the B5G/6G communication, and is committed to providing enlarged coverage, high speed and seamless connection, strong security and reliability. The SAGIN combines technologies from the ground, air and space networks, where the details are as follows.

\begin{itemize}
\item {\bf Ground-based network:} The ground-based network is the foundation of SAGIN, including elements such as ground-based Internet and mobile communication networks. To meet the communication requirements from cloud centers to individual mobile users, a variety of machine learning is adopted for optimization. {For example, in \cite{fu2022traffic}, a traditional machine learning method based on two-dimensional long short-term memory (LSTM) is proposed to predict network traffic and achieve resource allocation. However, due to the dynamic nature of ground-based networks (i.e., user mobility and traffic fluctuations), LSTM may encounter delays in adapting to network changes, thus preventing timely and optimal resource allocation.}

\item {\bf Air-based network:} The air-based network establishes high-altitude communication platforms and UAV self-organized networks. The high-altitude platforms provide a wide range of communication services, in which UAV networks composed of multiple UAVs offer autonomous connections to enhance communication support for ground users. Several studies apply traditional machine learning-based methods to air-based networks. {For example, in \cite{zeng2021simultaneous}, a deep reinforcement learning (DRL) approach was proposed to optimize UAV trajectories to minimize the weighted sum of execution time and expected transmission interruption duration. However, traditional DRL requires a high computing resource consumption, a large amount of calculation, and a long convergence time, which bring challenges to real-time decision-making in highly-dynamic aerial environments.}

\item {\bf Space-based network:} The space-based network is composed of a variety of satellite systems as backbone networks that ensure global connectivity and facilitate specialized communications. To achieve the goals of wide coverage, high bandwidth and low latency,  machine learning-based methods were developed for the space-based networks. {For example, in \cite{karavolos2022satellite}, a cooperative transmission scheme for satellites and airborne base stations was developed using K-means and K-medoids. However, due to the sensitivity of K-means methods to the initial centroid location and their two-dimensional assumption of clusters, they may have difficulty in fully covering complex channel data of three-dimensional spatial communication. This is especially in situations such as satellite signal interference or multi-path propagation.}
\end{itemize}

Therefore, to achieve better design and optimization, there is an urgent need to introduce entirely new machine learning algorithms, especially generative AI.

\subsection{Overview of generative AI} 
Generative AI is a distinguished subdomain of AI, exmphasizing on the conceptualization and generation of content. It emerges from the aspiration to enable machines to fabricate innovative and unprecedented data that faithfully mirrors the inherent patterns, structures, and nuances encapsulated in the training datasets. Generative  AI integrates a plethora of sophisticated models, including VAE, GAN, GDM, and transformer-based model (TBM), each espousing unique methodologies for assimilating and generating data.

\begin{itemize}
\item {\bf Variational Autoencoders (VAEs):} VAEs are generative models that use variational inference to model high-dimensional data, allowing them to generate new, consistent data instances. In SAGINs, VAEs facilitate the modeling and generation of adaptive communication protocols and diverse network configurations. For instance,  in \cite{bano2023generative}, a generative framework based on VAE was proposed to generate synthetic data for LOS estimation in non-terrestrial networks. In this case, the generated synthetic data does not contain identification information from the original dataset, allowing for its public use without violating privacy.

\item {\bf Generative Adversarial Networks (GANs):} GANs are based on a game-theoretic approach involving two parts, i.e., a generator and a discriminator， which are effective in creating high-quality data. In SAGINs, GANs are used to mimic real network traffic patterns, aiding in better network traffic management and anomaly detection. For example, in \cite{cai2023satellite}, a network traffic prediction method based on LSTM and GAN was proposed to address the significant variations in service traffic access across different regions of satellite networks. Here, GAN was utilized to augment the dataset and prevent the occurrence of training overfitting issues.

\item {\bf Generative Diffusion Model (GDMs):} 
GDMs replicate random diffusion processes to map data structures, leading to the creation of realistic data instances. In SAGINs, generative diffusion models are instrumental for simulating diverse network states and communication channel conditions.  For example,  in \cite{awasthi2023anomaly}, a GDM detection algorithm was proposed that can capture fast-moving anomalies in satellite data without any moving components.  Simulation results showed that the proposed GDM detection algorithm outperforms other baseline methods.

\item {\bf Transformer-Based Models (TBMs):} TBMs are characterized by a self-attention mechanism that can manage data, learn complex dependencies, and generate coherent contextual instances. In SAGINs, TBM is  used to simplify sequential communication protocols and ensure consistent exchanges between different network domains. For example, in \cite{ding2022bi}, a model based on TBM was proposed for semantic segmentation of remote sensing images in satellite networks, where an adaptive fusion module was designed to blend semantic information between features of different scales through a self-attention mechanism.
\end{itemize}

The  features of  generative AI models are summarized in Table~\ref{table_generative AI}. {Moreover, note that deep reinforcement learning (DRL) focuses on learning suitable policies through environmental interactions, primarily emphasizing decision-making and control strategies. In contrast, generative AI is designed to generate new data points, strategies, or content based on existing data, with a strong focus on data augmentation and scenario simulation. Although some related work has combined these two approaches \cite{du2024enhancing}, generative AI complements RL by providing enriched data and potential strategies that enhance RL's decision-making capabilities, particularly in complex and dynamic environments.}

\begin{figure*}[!t]
\centering
\includegraphics[width=\textwidth]{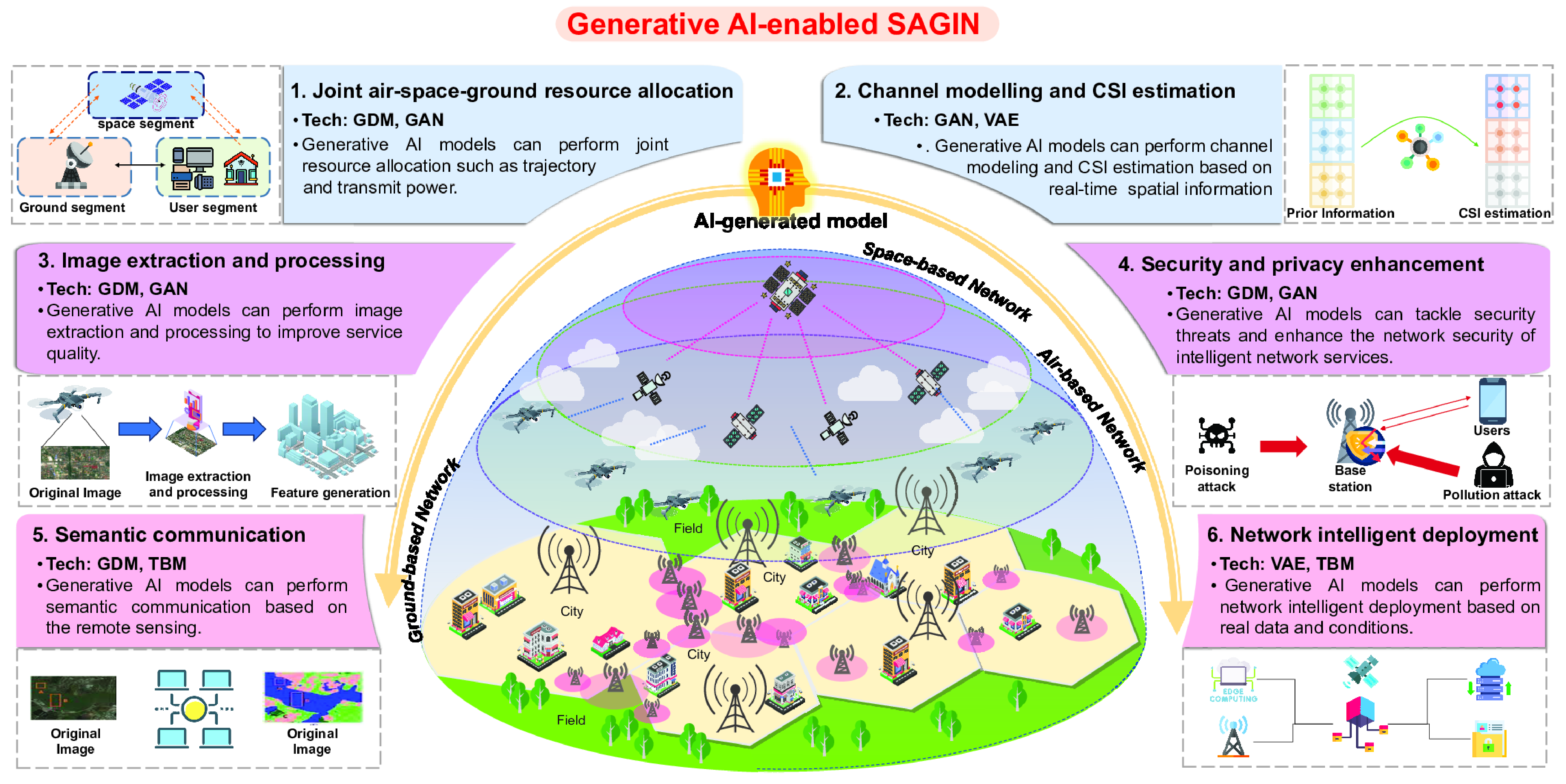}

\caption{{Generative AI-enabled SAGIN. Different generative AI technologies can be used to perform channel modeling and CSI estimation, joint air-space-ground resource allocation, intelligent network deployment, semantic communications, image extraction and processing, and security and privacy enhancement.} }
\label{fig_PSNC_1}
\end{figure*}

\section{Generative AI-enabled SAGIN }

In this section, we explore the issues and solutions of generative AI in SAGIN. Next, we discuss the  challenges associated with implementing generative AI-enabled SAGIN.

\subsection{Research Issues and Solutions.}

As shown in Fig.~\ref{fig_PSNC_1}, generative AI-enabled SAGIN can enhance channel estimation, optimize resource allocation, and ensure intelligent networks by incorporating these advanced generative AI technologies.  {For generative AI, it can be primarily implemented in the cloud. The cloud provides substantial computational power and storage capabilities, which are essential for running generative AI models. By deploying these models in the cloud, we can leverage these resources to handle the intensive computations required for training and inference. Given SAGIN's diverse and distributed components, such as satellites, UAVs, and ground-based transmitters, the cloud-based implementation allows for centralized processing and management of the AI models. Once trained, the generative AI models can generate strategies and data that can be used by these SAGIN components. This approach is reasonable because it allows the generative AI module to utilize the extensive computational resources of the cloud while distributing the results to the end or edge for application.} Next, we summarize major research issues and solutions of generative AI-enabled SAGIN.

\subsubsection{\bf Channel modeling and CSI Estimation}
{Traditional channel modeling and CSI estimation often rely on predefined models based on historical data and empirical measurements, necessitating manual updates and revisions to account for the dynamic SAGIN environment. Iterative methods, such as the RL or Kalman filters, enhance estimation accuracy through repetition and feedback. However, they lack an ability to autonomously adapt to unmodeled channel conditions or recognize complex patterns in data.  In contrast,} generative AI, specifically  GAN and VAE, provides an  approach to enhance channel modeling and CSI estimation in SAGIN through using real-time spatial information and analyzing radio wave propagation. Specifically, GAN, with its two-part neural architecture, effectively simulates channel perturbations by creating virtual channel scenarios. Its generator introduces underlying channel conditions, while the discriminator distinguishes between real and simulated conditions. This iterative process enables GAN to develop reliable channel models that reflect real-world SAGIN conditions \cite{9169908}. Moreover, VAE can learn probabilistic mappings and is used for CSI estimation from observed data. Through the compression and reconstruction of wireless channel data, VAE is able to provide accurate CSI predictions, aligning empirical data with theoretical models.


\subsubsection{\bf Joint Air-Space-Ground Resource Allocation}
{Traditional joint resource allocation often utilizes algorithms like linear programming or convex optimization, which are based on well-defined mathematical models. These methods, while precise for known scenarios, might not adapt well to the dynamic changes in SAGIN environments. Techniques such as fixed power allocation and static trajectory planning are employed, but they may not guarantee optimal performance under varying conditions. In contrast,} generative AI, specifically GDM and GAN, provides an approach for joint resource allocation in SAGIN. These generative AIs are capable of optimizing trajectories of aerial devices and dynamically adjusting transmit power to meet the various requirements of air and ground communications.  Specifically, GDM can capture patterns such as time series dependence and spatial correlation to model channel distribution in the time and spatial domains, respectively, which can provide a view of an entire communication environment. With this capability, GDM can effectively identify suitable trajectories and power levels, ensuring efficient communications and minimal interference \cite{10172151}.  Moreover, in the GAN structure, the generator produces hypothetical resource allocation scenarios, and the discriminator assesses and refines the efficiency of these allocations, enhancing resource utilization and network performance.

\subsubsection{\bf Intelligent Network Deployment}
{SAGIN network deployment has been guided by grid-based methodologies, genetic algorithm, and RL. These techniques, though effective in specific settings, might not be optimal for the diverse and dynamic demands of SAGIN environments. Moreover, they often require manual intervention or adjustment.  In contrast,} generative AI, specifically VAE and TBM, provides an approach for intelligent network deployment in SAGIN. Specifically, VAE can encode high-dimensional data into a lower-dimensional representation and decode it. This generative capability allows VAE to discern underlying patterns and dependencies, leading to efficient and adaptive network deployment. Furthermore, with its self-attention mechanism, TBM ensures that all SAGIN users (i.e., ground users and air users) can interact with any other user. This can improve the heterogeneity of the SAGIN network and enhance network decisions about topology, node placement, and resource allocation\cite{9943807}.


\subsubsection{\bf Semantic Communications}
{Traditional semantic communications rely on methods like ontology-based data representation or convolutional neural network (CNN). While they provide structured ways to represent and communicate data, they can become rigid and unable to adapt to new or evolving data patterns in SAGIN. In contrast,} generative AI, specifically GDM and TBM, provides an effective semantic communications approach for SAGIN based on the remote sensing. Specifically,  thanks to its ability to simulate data distribution, GDM is able to understand the semantic structure of communication datasets. Therefore, GDM can obtain valuable information from a wide range of datasets to facilitate semantic communications. On the other hand, through its self-attention mechanism, TBM is able to establish a context-aware communication environment, which ensures that key information is always maintained in the SAGIN network. For example, when satellites in SAGIN make observations about channel condition or remote sensing, TBM helps identify relationships between geographic features such as vegetation, water, and urban areas, which maintains the fidelity of the relationships during data transmission and interpretation.

\subsubsection{\bf Image Extraction and Processing}
{Traditional image extraction and processing in SAGIN have often been performed using techniques like edge detection, fourier transforms, and CNN. While these methods can process images, they might not be adept at handling diverse range and complexity of images in SAGIN scenarios. In contrast,} generative AI, specifically VAE and GAN, provides an approach for image extraction and processing  in SAGIN, which paves the way for diverse applications such as autonomous vehicle navigation with Internet of Vehicles (IoV). Specifically, VAE can effectively compress and extract high-dimensional image data and reconstruct it through generating data, making it applicable to a variety of SAGIN datasets \cite{10172151}. Alternatively, through the generator-discriminator structure, GAN is able to  improve image quality by generating enhanced synthetic data. The generator generates high-quality versions of images while the discriminator distinguishes them from the original dataset, thereby iteratively improving the image output.

\subsubsection{\bf Security and Privacy Enhancement}
{Traditional security measures in SAGIN primarily utilize intrusion detection systems (IDS), and deep belief network (DBF). These rule-based systems, while essential, may not always identify emerging  or evolving threats, especially those designed to exploit unknown vulnerabilities. In contrast,} generative AI, specifically GDM and GAN, provides an approach to enhance security and privacy in SAGIN.  Specifically, thanks to its ability to model data distribution, GDM is able to identify inconsistencies in typical data flows, ultimately enabling it to flag irregular network behavior or security vulnerabilities. Moreover,  GAN composed of dual neural entities operates within security vulnerabilities assumed by the generator design, while the discriminator enhances its detection algorithm based on these simulated vulnerabilities. This iterative process of confrontation gradually enhances the network’s defense capabilities and ensures improved resiliency against current and emerging cyber threats \cite{9298135}.


\subsection{Challenges}


As generative AI is integrated into SAGIN, the system's intelligence and seamless communication can be improved. However, certain challenges still exist as follows.

\subsubsection{\bf Real-time Responsiveness}
Integrating generative AI into SAGIN demands immediate and responsive processing and decision-making capabilities. Ensuring real-time response is critical, as any delays can severely compromise the effectiveness of the fast-changing network. For example, suppose a method takes too long to process and update channel information, whether generative AI or discriminative AI algorithms. In that case, it may not be able to adapt quickly enough to dynamic environmental changes, leading to degraded service quality and increased latency. The ability to swiftly adapt to changing conditions, reduce latency, and improve overall network performance is crucial for successfully integrating generative AI into SAGIN.

\subsubsection{\bf Integration Overhead}
{The integration of generative AI into SAGIN introduces additional computational and communication complexities. Generative AI technologies, especially sophisticated ones like GANs and VAEs, can lead to increased computational resources and time overhead. Efficiently managing this integration overhead is essential to ensure that the enhanced capabilities do not come at the cost of system performance or network efficiency. For example, training a GAN model can require hundreds of GPU hours, necessitating significant cloud resources. Additionally, the real-time application of these models may introduce latency due to data transmission between the cloud and SAGIN components. This overhead can slow down the network's responsiveness and affect the system's overall performance.}

\subsubsection{\bf Scalability Issues}
{SAGIN's architecture includes ground-based, air-based, and space-based components, presenting scalability challenges when integrating with generative AI. The requirement for generative AI technologies will increase as the network grows and more devices and nodes are added. Ensuring that generative AI technologies can scale effectively without compromising their performance or network reliability is critical to the long-term success of generative AI-enabled SAGIN. For instance, as more UAVs and satellites are deployed, the computational demand for real-time data processing and decision-making increases, leading to significant energy consumption. The complexity of managing these additional resources can result in higher operational costs and potential sustainability issues. }

\section{Generative AI-enabled Channel Information Map Construction}
In this section, we propose a framework that utilizes a GDM-based model for construct channel information map to enhance quality of service.
\subsection{Motivations}
The SAGIN framework integrates different networks in a hierarchical structure. However, in densely urban landscapes, a significant issue is regarding the non-line-of-sight (NLOS) signals. In essence, while direct line-of-sight (LOS) signals originate from satellites, drones, or ground-based base stations, non-line-of-sight signals are subject to interference from urban obstructions such as tall buildings and giant bridges. This causes these signals, especially those from satellites and UAVs, to reflect multiple times before reaching their target. These multiple reflections introduce ``multipath errors", where the direct and reflected signals have different arrival times and phases \cite{li2023nlos}. This discrepancy results in notable position errors, making it challenging for applications that demand high accuracy of signal.

\begin{figure*}[!t]
\centering
\includegraphics[width=\textwidth]{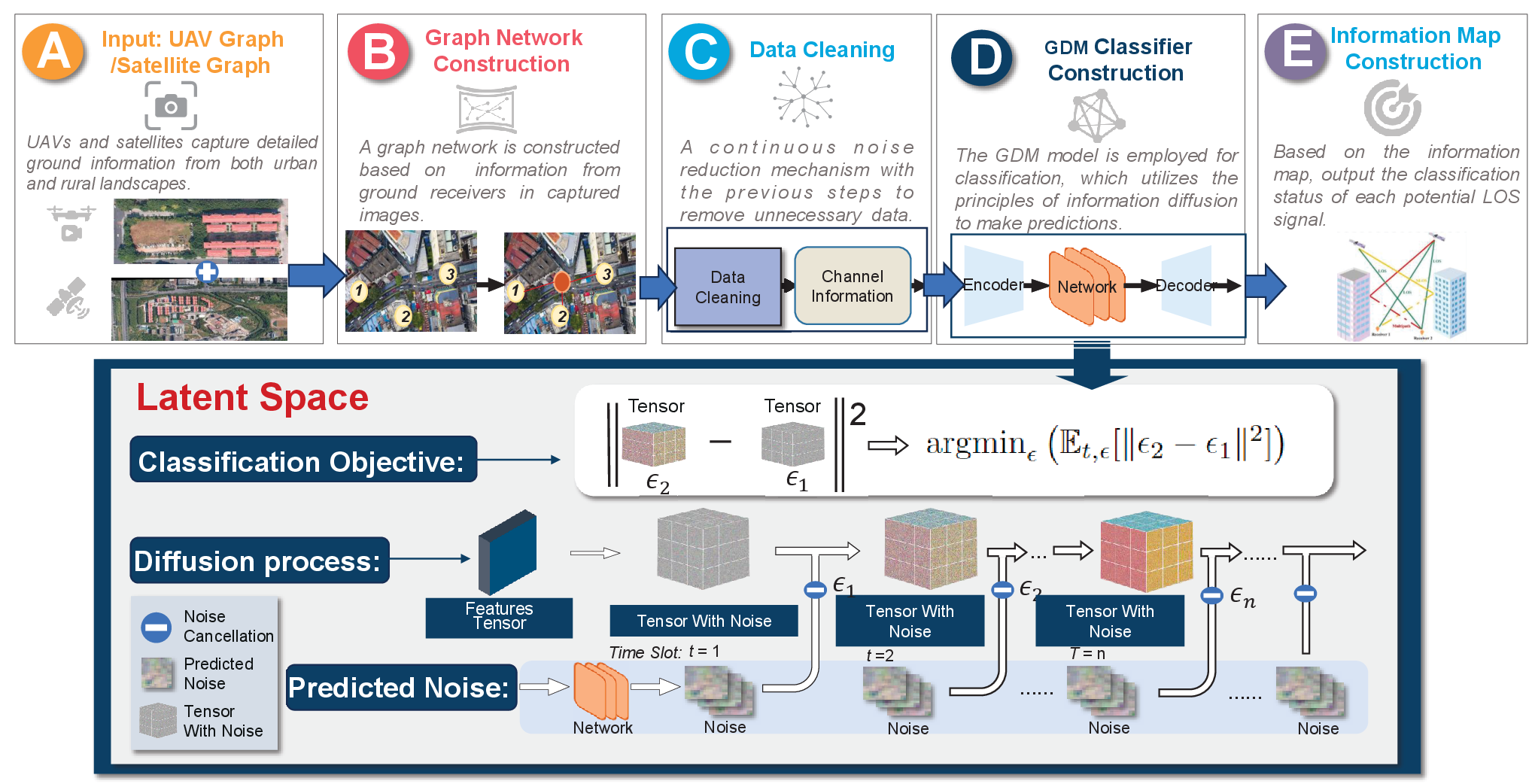}

\caption{A process on channel map architecture, where the GDM classification method is proposed. In GDM classification, latent space is used for capturing essential data features while ignoring non-essential variations；  Feature tensor denotes a multi-dimensional array encapsulating the significant attributes of the data.  Predicted Noise denotes the model estimation of structured noise to be eliminated from the data. Tensor With Noise denotes that data representation posts incremental noise addition over various time slots.}
\label{framework}
\end{figure*}
\subsection{Proposed Framework}

Addressing the implications and seeking solutions to the challenges posed by multipath errors in SAGIN are possible through harnessing the potential of generative AI technologies. Accordingly, as shown in Fig.~\ref{framework}, a GDM-based channel map framework is proposed.
{Note that while equalizers and MRC are highly effective in mitigating multipath errors and improving signal quality, they primarily rely on linear signal processing techniques. These methods are well-suited for certain environments but can struggle in complex scenarios where channel characteristics are highly non-linear and dynamic, such as dense urban areas with significant reflections and obstructions. Moreover, non-ML frameworks such as equalizers and MRC often require precise knowledge of the CSI and involve complex calculations. These methods can be less adaptable to rapidly changing conditions, as they are typically designed for specific scenarios and may not generalize well to diverse and evolving network environments.  In contrast, generative AI models, such as GDM, excel in capturing and modeling non-linear relationships within data. This makes them particularly suitable for handling multipath environments' intricate and variable nature. Generative AI models can learn from vast amounts of data and continuously adapt to changing conditions, providing more accurate and robust channel modeling and state estimation even in highly dynamic scenarios. In particular, GANs utilize latent representations to transform raw input data into a more abstract and flexible form \cite{9169908}, which is especially beneficial for complex signal-processing tasks. These latent representations allow GANs to capture the underlying structure and patterns of the data, leading to more effective and adaptive signal processing than traditional methods. This flexibility enables generative AI to provide more comprehensive solutions in scenarios where traditional linear techniques may fall short.} Next, our GDM-based channel map framework consists of the following functions.
\begin{itemize}
\item {\bf Step 1: Image Acquisition:} Within the scope of SAGIN, data collection is a continuous process. Satellites and UAVs are regularly deployed to capture detailed ground information from both urban and rural landscapes. While satellites play a major role in capturing overview ground information, UAVs are capable of providing more detailed local images. Once acquired, the information, e.g., captured images, are transferred to the processing centers, providing a foundation for subsequent steps of the framework.

\item{\bf Step 2: Image Processing:} At the processing center, attention should be paid to the division of areas from the image where the LOS signal is inferred. From these delineated areas, relevant data such as excess delay, signal paths, and transmit power of the transmitters (i.e., UAVs or satellites) are extracted through ground-based receivers. For a comprehensive representation of the indicators, this data is structured into a graphical network.

\item {\bf Step 3: Elimination of Redundant Information:} In SAGIN, especially under complex multipath conditions, there is always potential for noise or superfluous information. Specifically, during image processing, unintended anomalies can emerge from data redundancy, leading to an environment representation (EP) that might be cluttered with irrelevant details, hindering the accurate identification of the target transmitters. To solve this, a continuous noise reduction mechanism based on a GDM is initiated. This process refines the diffusion network by removing unnecessary noise, ensuring that the resulting classification is relevant to the current target.

\item{\bf Step 4: Zero-shot Classification:}
 Once the data is in place, stable diffusion density estimation\footnote{{Stable diffusion density is a probabilistic modeling technique used to estimate the distribution of data points in a high-dimensional space. In our framework, this technique involves modeling the spread and concentration of data points to understand their underlying structure.} } is utilized. These estimates possess the capability to drive zero-shot classification\footnote{{Zero-shot classification is a method in machine learning where the model can correctly classify instances from classes that it has never encountered during training. In our framework, zero-shot classification enables the model to identify and categorize new types of channel conditions without requiring additional training data for those specific conditions.}}, removing the necessity for additional training. The classification is accomplished by discerning patterns and variations within the network and aligning them with density estimates.  

\item {\bf Step 5: Channel Information Construction:} 
Finally, the channel information map is constructed to classify each potential LOS signal to improve the communication and service quality of the entire system.
\end{itemize}


\begin{figure*}[!t]
\centering
\includegraphics[width=0.8\textwidth]{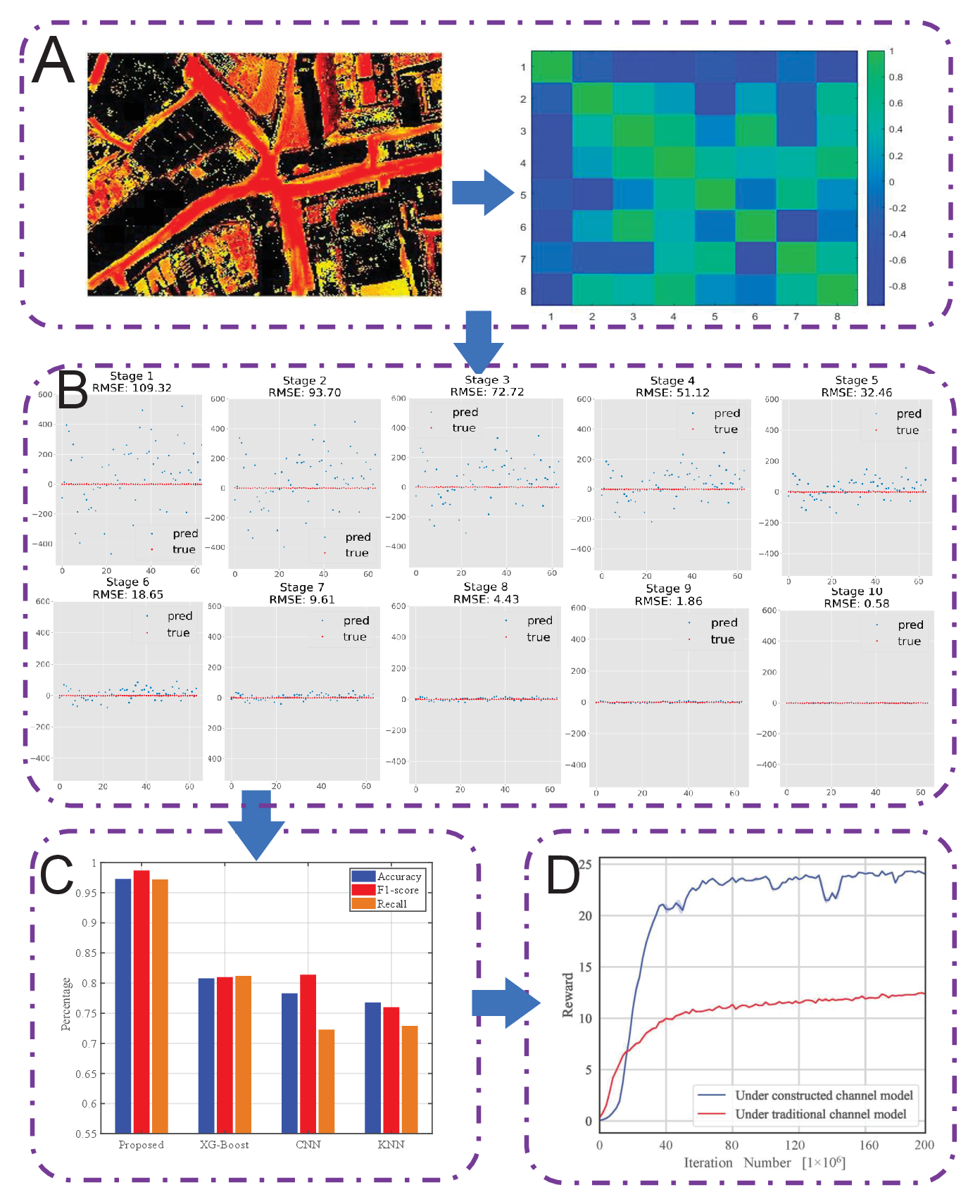}
\caption{{The simulation results, where the Part A shows about the attention map, Part B shows the GDM classification process, Part C verifies the validity of the classification, and the Part D shows the application of using the channel map.}}
\label{simulation}
\end{figure*}

\subsection{Case study}

As mentioned above, using images taken by UAVs and satellites, ground receivers are first used to extract corresponding picture information, and then map construction and channel classification are performed. {{Here, the channel map is generated at a cloud-based processing center. The cloud provides substantial computational power and storage capabilities essential for running sophisticated generative AI models like GDM. In our framework, data collection is a continuous process facilitated by satellites and UAVs. These devices capture detailed ground information, which is then transferred to the cloud for processing. The use of ground-based sensor nodes further enhances the accuracy and richness of the data collected. In particular, the collected images are preprocessed to identify and segment areas where the LOS signals are inferred. This segmentation process helps in isolating the relevant data needed for further analysis. The data is then subjected to noise reduction using the GDM, eliminating redundant information and focusing on the essential details. Stable diffusion density estimation is employed to classify the data, removing the necessity for additional training. Finally, the channel information map is constructed, categorizing each potential LOS signal, thereby improving the communication and service quality of the entire system. Here, SAGIN plays two critical roles, i.e., capturing image information and using this information to extract spectrum map features that enhance the overall network performance. This creates a beneficial cycle where the collected data improves the network, and the improved network, in turn, facilitates better data collection.}} {The simulation involves a single satellite, one UAV, and a ground-based access point. This setup includes four users within the simulation environment.  Backhaul is performed using GNSS signals to ensure precise location data and synchronization. Resource allocation is managed by classifying the collected data using a GDM, which processes and categorizes the information to optimize network resources. The dataset utilized for this simulation is a classic GNSS type, comprising raw measurements from the f9p receiver. This dataset includes elements, such as GPS time, satellite code, code and phase measurements in L1 and L2 bands, Doppler measurements, signal strength, LOS/NLOS labels, pseudorange error, and the ground truth location (latitude, longitude, and height). For attention map,  we use 100,000 images, of which approximately 75\% are used for training and 25\% for validation.} {In order for the ground receiver to more accurately extract relevant data such as the direction angle of the transmitter (i.e., UAVs or satellites), the key semantic information in the images is first extracted,} as shown in part A of Fig.~\ref{simulation}. For the left results, the {attention map} reflects the importance of map information through different brightness levels, where the more brightness indicates the greater weight of the corresponding map information. For instance, the information weight of the road area and roof area is greater because there is no obstruction between them and the link information is better. Moreover, the {attention map} depicted on the right uses a color scale from blue to green. The dark blue represents the areas with the least attention, while the bright green represents the areas with the most attention.  By comparing the two results, it is evident that areas with more brightness in the original image usually correspond to areas with brighter green in the {attention map}. This demonstrates the system's ability to assign higher weights to specific parts of the image and focus more on areas such as roads and rooftops, as these areas provide a clearer LOS.

After obtaining the channel data information, it is required to be classified. In this step, the GDM model is employed for classification, which utilizes the principles of information diffusion. The process of diffusion classification is shown in Part B of Fig.~\ref{simulation}.  Fig.~\ref{simulation} (B) shows the stage scatter plot as training time changes. Each point is a set of training data, and the root mean square error (RMSE) is used as the evaluation target. From the first to the last image, one can clearly see the RMSE gradually decreasing, from the starting value of 109.32 to only 0.58. This indicates that the accuracy of model's predictions significantly improves as the diffusion process proceeds. Additionally, each point on the scatter plot represents the model prediction error. At the beginning of the diffusion, these error points are widely distributed and far away from the red baseline (i.e., 0 represents the LOS signal). However, as the diffusion steps proceed, these points gradually converge toward the red baseline, which further verifies the improvement in model accuracy.

Next, to verify the effectiveness of the proposed GDM classification, we compare it with several commonly used classification algorithms such as the K-nearest neighbors (KNN) method, extreme gradient boosting (XG-boost) method, and CNN method, with the results shown in part C of Fig.~\ref{simulation}. The results show that the proposed GDM classification method is superior to the other benchmarks regarding accuracy, F1 score, and recall rate. This is because the GDM model can effectively capture the distribution and complex patterns of the underlying data for high-latitude or complex data. Unlike traditional classification methods, which may be susceptible to noise or data irregularities, GDM utilizes a diffusion process to gradually transform the data, ensuring a smoother, more robust representation. This inherent ability to extract complex data structures allows it to perform better when other algorithms struggle with overfitting or misclassification. It is worth noting the performance specifics of the three benchmark algorithms. The KNN method, although intuitive, sometimes has difficulty handling high-dimensional data and is sensitive to the choice of distance metric. The XG-boost method is a gradient enhancement method but is prone to overfitting. The CNN method works on the principle of local receptive fields and hierarchical feature extraction, which may not always be optimal for data sets that lack spatial structure. While each method has its merits and applications, its limitations in certain scenarios make the GDM's inherent ability to extract complex data structures even more commendable. {Moreover, using the GDM-based model for channel information mapping leads to faster updates in network configurations. Specifically, it shows that the GDM-based approach can achieve a reduced RMSE within 10 stages. In contrast, traditional methods such as CNN and K-means required over 100 iteration stages to reach similar accuracy levels. This significant reduction in the number of iterations needed for convergence demonstrates the efficiency of the GDM-based model, ensuring that the network can respond swiftly to dynamic environments, maintaining high performance and reliability.}

Finally, based on the classification results, the channel map is constructed. To further verify the effectiveness of the constructed channel model, part D of Fig.~\ref{simulation} shows the cumulative reward value obtained with the increase of the number of iterations using the same PPO-based strategy for different types of channel models. Here, the corresponding curves are smoothed via a sliding window to provide a clearer overall trend of the raw results. The results show that regardless of the types of channel models, the RL-based strategy can converge with an increment of iterations. {The drop in the reward after around 120 iterations can be attributed to the exploration mechanism inherent in RL algorithms, specifically the PPO-based method used in our study. During the training process, the RL agent occasionally explores less optimal actions to improve its policy and avoid local minima. This exploration phase can temporarily reduce the cumulative reward as the agent evaluates various actions. However, as training progresses, the agent learns from these explorations and adjusts its policy to achieve higher performance, leading to an increase in the reward again.} Furthermore, the channel model constructed using our method can achieve higher performance gains. This is because the channel constructed by our method more accurately reflects the channel conditions under better classification conditions, allowing related resource optimization methods to find more reasonable and feasible solutions, i.e., the suitable transmit power.

\section{Future Directions}
In this section, we outline three main future directions on implementation and enhancement for generative AI-enabled SAGIN scenario. 

\subsection{Multi-modal Integration for Holistic Connectivity}
SAGIN integrates various data sources, such as images, videos, audio, and radar scans, from satellites to ground sensors, providing information. Generative AI can seamlessly integrate this multi-modal data to view the SAGIN operating environment comprehensively. By merging data from different modalities, SAGIN can generate a detailed environmental model, enhancing the network's ability to make informed decisions. For example, combining radar scans with satellite images can improve object detection and tracking accuracy, while integrating audio data can enhance situational awareness in emergency response scenarios. Technical approaches like deep learning-based data fusion and multi-modal neural networks can be employed to combine and analyze these diverse data sources effectively. This multi-modal integration enriches the contextual intelligence available to the network, enabling more precise and effective responses to dynamic conditions and improving overall connectivity and performance.

\subsection{Sustainable and Scalable Operations}
{With growing concerns about energy consumption and environmental impact, it is crucial for SAGIN to operate efficiently and sustainably. Non-terrestrial networks, such as those involving satellites and high-altitude platforms, generally have lower energy consumption than terrestrial networks. Generative AI can enhance these efficiencies by optimizing trajectory planning and communication strategies. For example, generative AI can analyze historical data and current operational metrics to predict future network demands, recommending optimal satellite trajectories that minimize fuel consumption. Additionally, generative AI can optimize communication protocols to reduce energy usage during data transmission and processing. By continuously learning and adapting to environmental conditions and network demands, generative AI ensures that SAGIN operates sustainably while maintaining high performance. }

\subsection{Support for AI-Generated Content Services in SAGIN}
{Incorporating AI-generated content (AIGC) services into SAGIN emphasizes the potential for significant technological advancements. Generative AIGC services can change how data is generated, processed, and utilized across the network. For instance, AIGC services can create personalized content such as real-time traffic updates, emergency alerts, and customized multimedia content, enhancing user engagement and satisfaction. By leveraging AIGC, SAGIN can provide dynamic, responsive, and context-aware services that adapt to real-time user needs and environmental changes. However, integrating these services introduces challenges related to computational and communication complexities. Efficiently managing these complexities to minimize integration overhead while maximizing the benefits of AIGC services is critical. Approaches like model compression, edge computing, and efficient data transmission protocols can help address these challenges.}

\section{Conclusion}
In this article, we explored the integration of generative AI and SAGIN. Specifically, we first reviewed SAIGN, including ground, air, and space-based networks. Then, generative AI technologies, such as GAN, VAE, GDM, and TBM were introduced. Next, certain issues and solutions of generative AI-enabled SAGIN and potential challenges were presented.  {To enhance the quality of service for SAGIN}, we proposed a framework that utilizes a GDM-based model for constructing channel information map. The simulation results verified the effectiveness of the proposed framework. Finally, future directions regarding generative AI-enabled SAGIN scenarios were summarized.

\bibliographystyle{IEEEtran}
\bibliography{mylib}

\end{CJK}
\end{document}